\def\la{\mathrel{\hbox{\rlap{\hbox{\lower4pt\hbox{$\sim$}}}\hbox{$<$}}}}
\def\ga{\mathrel{\hbox{\rlap{\hbox{\lower4pt\hbox{$\sim$}}}\hbox{$>$}}}}

\def\msunyr{M$_{\odot}$ yr$^{-1}$}

\documentclass{kapproc} 

\usepackage[dvips]{graphicx}

\upperandlowercase

\normallatexbib 

\begin{document}

\articletitle{Conference Summary: Starbursts and Galaxy Evolution}


\author{Robert W. O'Connell}
\affil{Astronomy Department\\
University of Virginia\\
Charlottesville, VA \ \ USA 22903-0818}
\email{rwo@virginia.edu}

\begin{abstract}

Starbursts are extreme concentrations of star-forming activity with
mass conversion surface densities reaching over 1000 times higher than
normal for disk galaxies.  They are responsible for a large fraction
of all cosmic star formation.  They have shaped the cosmic landscape
not just in individual galaxies but though the effects of 
``superwinds'' that enrich the intergalactic gas, constrain
supermassive black hole growth, and perhaps facilitate cosmic
reionization.  This conference provided vivid testimony to the
importance of starbursts in galaxy evolution and to the speed with
which our understanding of them is being transformed by a flood of new
pan-spectral data, especially at high redshifts.  A key result is that
it appears possible to scale the physics of smaller star-forming events
upward to encompass starbursts.

\end{abstract}

\section{Introduction: Solved \& Unsolved Problems}

\noindent {\it ``Galaxy formation is a solved problem.''} That was the
most memorable quote from the last meeting I attended at Cambridge,
the 1996 conference on the Hubble Deep Field.  The prominent
astronomer who offered this sentiment was perhaps a little premature,
but his excess of enthusiasm was forgivable considering the stream of
beautiful data on the high redshift universe that had just begun to
emerge.  That stream has exponentiated over the last 8 years, and it
may well be that we can solve the problem of galaxy formation within
the next couple of decades.

Before that is possible, however, we need to understand star
formation.  We have already solved the basic problem of stellar
structure and evolution.  That can fairly be said to be the primary
accomplishment of astrophysics in the 20$^{th}$ century (especially
because it had been several million years since humans had first
wondered about the stars!).  There are only a few remaining dark
corners of the evolutionary process.  But one is crucial: 
star formation.  This is central to galaxy astrophysics, but
the deficiencies in our understanding are obvious.
For instance, faulty prescriptions for star formation are
thought to be the culprits in discrepancies between predictions
of CDM models for galaxy formation and the observations.

A great deal of observational firepower will be directed at the
problem of star formation in the coming years, but we already know one
essential fact:  star formation is a {\it collective} process.  Most
stars (perhaps nearly 100\% in our Galaxy) form in clusters.  There
are strong interactions with the surrounding environment and among
protostars.   Quantifying feedback processes, both positive and
negative, is a key to understanding star formation.  All this means
that star formation is a more difficult problem than was the
astrophysics of isolated stars.  Progress will be importantly informed
by observations of other galaxies and a wider range of environments
than are found in our Galaxy.

Starbursts are important because they are clearly a collective
phenomenon and represent one extreme of the star formation process.
Furthermore, they are bright enough to be detected throughout the
observable universe and can serve as tracers of the cosmic history of
star formation.  This conference provided vivid testimony to the
importance of starbursts as keys to galaxy evolution and to the speed
with which our understanding of them is being transformed by the
flood of new data, especially at high redshifts.

\section{What Are Starbursts?}

The term {\it starburst} conveys the dual notions of intensity and
limited duration.  There is no strict definition, however, so the term
has been used (or abused) to encompass a huge variety of star
formation events.  Several speakers proposed useful definitions, and I
will follow their lead.  

Significant star formation is a hallmark of about half the galaxies in
the local universe.  There are several convenient proxies for active
star formation:  blue optical/UV colors, emission lines,
or strong infrared output. Although there was a general awareness of
the statistics, the Sloan Digital Sky Survey has recently brought home
the unmistakable {\it bimodality} of optical colors:  galaxies fall
into either a red or blue sequence (separated by about 0.4 mag in B--V
color), with few systems in between.  This means they are either
active star formers (blue) or have not hosted significant star
formation for $\ga 1$ Gyr.  Interphase types are rare because once
star formation ceases, color evolution from the blue to the red
sequences occurs in only $\sim 500$ Myr (as long as the active
galaxy is itself old).  

The blue systems are mainly disk-dominated.  The normal
structure and dynamics of disks favor relatively slow conversion of
gas to stars.  Global {\it self-regulation} within disks is evidently
effective over long timescales because there are good correlations
between ionizing populations (lifetimes $\la 10$ Myr) and broadband
optical colors (characteristic times of $\ga 1000$ Myr).  The range of
what might be called ``quiescent'' star formation
encountered along the normal Hubble sequence is about 4 orders of magnitude
in both star formation rate ($\dot s$, measured in \msunyr) and star
formation surface density ($\Sigma_{SFR}$, measured in \msunyr\ kpc$^{-2}$).
A significant number of local galaxies, mostly dwarfs, exhibit elevated
activity, ranging above $\Sigma_{SFR} \sim 0.1$, which
might be called ``enthusiastic'' star formation.  This accounts for
only $\sim$ 15\% of all local star formation.  

The most interesting cases, naturally, are the ``psychopathic'' ones at
the extremes, which are the starbursts.  These are often, though not
always, associated with a large disturbance to normal disk
kinematics.  The central feature of a starburst is the concentration
of star-forming activity and especially the large feedback it produces
on its surroundings, often driving a ``superwind'' out of the host
galaxy.  For definiteness, I will define a ``major starburst'' as an
episode where such effects are important.  This requires $\Sigma_{SFR}
> 1$
\msunyr\ kpc$^{-2}$, equivalent to $\Sigma_{L(BOL)} > 10^{10}$
L$_{\odot}$ kpc$^{-2}$ or about $1000\times$ normal values for disk
galaxies.  The relevant quantities must be averaged over a finite cell
size in area and time (say 1 kpc$^2$, 10 Myr), and the initial mass
function (IMF) must contain massive stars capable of producing
ionization, winds, and core-collapse supernovae. 

The fact that this definition is arbitrary is actually significant:
the properties of star formation regions appear to be {\it continuous}
across the range of amplitudes observed.  

Starbursts constitute an important fraction of all detected
high-redshift galaxies.  Because of the fierce dimming effects in
distance modulus and surface brightness at large $z$, there is a
powerful selection effect operating here.  Nonetheless, statistics
based on co-moving volume densities have shown that while major
starbursts are rare locally ($\sim$ 0.5\% of nearby systems) they were
much more common at earlier times (perhaps 15\% in number and
70$\times$ in luminosity density at $z = 1$).  IR-bright starbursts
have been responsible for as much as 80\% of the local stellar mass.

Although we don't have a definitive understanding of starbursts and
their effects yet, the wealth of new data is making progress rapid.
Along the way, there are two central difficulties: (1) major
starbursts are rare in the local universe, and we are forced to {\it
scale up} our understanding of physical processes from local samples
and conditions; and (2) starbursts are notoriously complex 3-D
systems, made especially difficult to probe by often severe
differential internal extinction.

\section{Are Starbursts Important?}

Starbursts are certainly fascinating, but how important are they?  An
interesting way to frame the question is to ask: if we didn't know
that major starbursts existed from direct observations, would we have
difficulties explaining what we see in the universe?  That is, are
starbursts a {\it necessary inference} from other phenomena?  The
answer is an emphatic ``yes,'' and here is a tentative and incomplete
list of the essential fingerprints of starbursts based on issues
raised in this conference:

\begin{itemize}
\vspace{-0.15cm}

\item{Super star clusters:} These very massive but compact systems,
ranging from very young clusters still in dust cocoons to classic
globular clusters, can evidently form only in a
high pressure medium, with $P/k \sim 10^{8-9}$, over $10^4 \times$ 
higher than normal for disk galaxies.  This requires abnormal,
non-equilibrium conditions such as prevail in starbursts.  Young SSC's
are found to be mainly associated with interaction-induced starbursts.

\item{Massive bulges and E galaxies:} The high stellar densities found
in the centers of nearby early type galaxies imply conversion of large
amounts of gas to stars at rates equivalent to the most extreme
starbursts known, $\dot s \sim 1000$ \msunyr, if only a single event
was involved.  There is good statistical evidence that many, if not
all, ETG's originate from gas-rich mergers, which are well known
to produce violent starbursts.  A few large mergers rather than a
series of minor mergers are favored.  (The best direct evidence for
bulge-producing starbursts at early times is probably the high
redshift sub-mm sample---e.g.\ from SCUBA---with $\dot s$ up to $\sim
1000$ \msunyr.)

\item{Cosmic mass deposition in stars:} For a decade, we have been
able to estimate the conversion rate of gas into long-lived stellar
mass at redshifts $z \la 4$.  The present-day mean mass density cannot
by itself place very good constraints on the range of $\Sigma_{SFR}$
since over 13 Gyr have elapsed since the big bang (though, as noted
above, direct statistics on distant starburst progenitors can do so).
However, recent deep probes to $z \sim 2$, such as GDDS, reveal that a
considerable fraction of all stars then are in ``dead and red''
systems comparable to local gE galaxies, with little star formation in
the preceding 1.5 Gyr.  The data imply that massive galaxy assembly
begins early ($z_f \ga 3$--5) in dense regions and, since there is so
little time for this to happen, massive starbursts with $\dot s \ga
300$ \msunyr\ must be involved, possibly through gas-rich mergers. 

\end{itemize}

\noindent The remaining items on the list involve {\it superwinds}
generated by starbursts:

\begin{itemize}
\vspace{-0.15cm}

\item{Chemical enrichment of the ICM and IGM:} Although metal
abundances at higher redshifts are generally lower than prevailing
local values, much of the gas to the highest $z$'s yet probed has been
processed through stars.  The primordial generations of stars
responsible can be explored only theoretically now, but it is clear
that the natural mechanism for dispersing new metals from the halos in
which they form is a superwind.  

\item{The mass-metallicity relation in galaxies}:  The correlation
between larger galaxy masses and higher metal abundances has been
known for about 40 years.  Optical, UV, and IR observations are
currently providing much better information on metallicites and dust
abundances in both local and high-redshift samples. 
Again, the natural explanation for the mass dependence is that starburst
superwinds evacuate gas preferentially from lower mass systems.  

\item{Absence of super-supermassive black holes:} Supermassive black
holes (SMBH) are now thought to exist in all spheroidal systems, and
their masses are linked to the surrounding stellar population.  Their
growth by gas accretion is self-limited to the Eddington rate.
Nonetheless, in the absence of another inhibiting mechanism, SMBH
masses would exponentiate with an e-folding time of about 100 Myr.  
Star formation in nuclei is often found associated with SMBH's,
and it may be that starburst winds act to regulate SMBH
growth in the same way they limit star formation itself. 

\item{Cosmic reionization:} It is likely that massive stars, rather
than AGN's, are responsible for cosmic re-ionization at
redshifts $z \ga 7$.  But the optical depth in typical nearby galaxies
is such that only 3--10\% of ionizing photons can escape.  Superwinds
in protogalaxies may be necessary to clear out channels for ionizing
radiation.  

\end{itemize}

\section{We're All Pan-Spectral Now}

In the past 5 years, the necessity as well as the opportunities to attack
the problem of starbursts using a multi-wavelength approach have become
manifest.  No single band suffices, and the full EM spectrum from radio
to gamma-rays is now enthusiastically embraced in starburst research.
Some examples:  ({\it a}) stellar ages and abundances are best deduced
from UV-optical-nearIR observations; 
({\it b}) the best $\dot s$ estimator is $L(UV)+L(IR)$, meaning that
different instruments are always necessary;
({\it c}) a long-wavelength baseline is essential to overcome 
distortions by extinction of statistical samples and of physical inferences from
any given band;
({\it d})  starburst
regions can be opaque even at mid-IR wavelengths; radio/mm observations are
needed for the youngest ($\sim 2$ Myr) embedded sources;
({\it e}) mid-IR photometry and spectroscopy, now just coming into their
own with the Spitzer Space Telescope, show great promise as dust/gas
tracers within starbursts.

Understanding the physical coupling mechanisms between wavelength
domains is essential:  
({\it a})  there has recently been good progress in modeling the UV through
IR spectral energy distributions of starbursts taking all three major
components (stars, gas, dust) into account,  but this remains a key area
for additional effort; 
({\it b}) the long-recognized relation between radio continuum 
and far-IR dust emission in star-forming systems is sometimes said to be
the best correlation known in extragalactic astronomy, yet we 
do not fully understand its origins or implications for the star formation
process.  A less well established radio/X-ray correlation in young
systems is also important to understand.  

The fastest increments in observational insight into starbursts are
currently coming from Spitzer and GALEX (IR and UV).  Probably
the fastest increments for the coming decade will be from ALMA
(mm-wave).  

Although this conference emphasized observations, we cannot forget that
theoretical and computational astrophysics have to be part of a
``pan''-discipline approach to starbursts.

\section{The Limits of Spatial Resolution}

A hard lesson in the study of starbursts has been that their scales
and complexities push the limits of instrumental spatial resolution
even in nearby systems.  For instance, it is difficult: 
({\it a}) to measure the diameters of SSC's
($\sim$ 2-10 pc) in order to obtain reliable mass and IMF inferences;
({\it b}) to study superwind substructures in nearby starbursts
and to determine host morphologies in distant ones; and
({\it c})  to obtain kinematics of starbursts on the appropriate
physical scales.  

The {\it Hubble} Space Telescope has been the mainstay of high
resolution ($\sim 0.05^{\prime\prime}$) imaging and spectroscopy for 14
years.  An informal count shows that over half the contributions in
this conference relied in some way on HST data.  But HST will not last
much more than another 6 years even if NASA can find a safe way of
servicing it.  In the foreseeable future, we will have the EVLA and
ALMA for high-resolution radio/mm observations and JWST and
ground-based AO systems for high resolution near and mid-IR
observations.  However, it is doubtful that AO systems will operate
well for $\lambda \la 1 \mu$.  Unfortunately, there are no current plans
to replace or improve (to $\sim 0.01^{\prime\prime}$?) high resolution
optical/UV capability in space.  It is vital to remedy that situation.

\section{Scalability}

``Scalability'' was a major theme of the conference.  It arises from
two main concerns:  To what extent can we scale local starburst
systems to cosmically distant ones?  And to what extent can we scale
the physics of modest to extreme star formation amplitudes?  The
evidence, fortunately, is that scalability is {\it good}, implying
modest rather than fundamental adjustments with changes in environment
and scale.

The premier example of scalability is the Schmidt-Kennicutt ``law,''
under which $\Sigma_{SFR} \propto \Sigma_{GAS}^{1.4}$.  The quantities
refer to global averages over the surfaces of individual galaxies.  
The relation applies over a remarkable 6 decades.  As noted above,
a similar degree of scalability applies to the radio/far-IR
correlation for star forming systems.  

Other encouraging, if less firmly established, examples of scalability
include:

\begin{itemize}
\vspace{-0.15cm}

\item Congruences in EM spectral shape for starbursts across a wide range
of environments and amplitudes.  

\item The continuity of starburst properties across a large
range of amplitudes.  It is possible to define a scaling sequence between
the nearby (3.5 Mpc) archetypal starburst M82 ($L \sim 2\times10^{10} L_{\odot}$),
more distant ULIRGs ($10^{12} L_{\odot}$), and high redshift SCUBA
sub-mm sources ($10^{13} L_{\odot}$).

\item The smooth increase of starburst activity with lookback time
exhibits no evidence of a {\it transition} point where starbursts
suddenly become more important. 

\item Continuity of Lyman break galaxies (LBG's) at $z \ga 3$ with
more local systems.  Careful studies, lately including GALEX data,
show that properties (sizes, surface brightnesses, masses, kinematics)
of LBG's are continuous with those of lower redshift luminous blue
compact galaxies, some of which may be the progenitors of local dE
galaxies (i.e. dynamically hot systems).  

\item The mass-metallicity-extinction relation, which changes only slowly
with redshift and has no transition points.  The abundance scale seems to
decrease smoothly with redshift.  

\item The duration of starburst episodes is $\delta t \sim 100$ Myr and
seems similar at all redshifts.  Individual galaxies may experience a number
of such episodes.  

\item The IMF for star formation on the scales of star clusters or
galaxies now appears to be {\it universal} except in a small number of SSC's
where there may be changes in $M_{LOW}$.  The massive
star IMF appears universal, which is very important for analyzing
feedback processes.  (Progress here has been excellent despite many
complications, e.g.\ limited spatial resolution, large differential
extinction effects, and mass segregation.)

\item Star formation histories of nearby galaxies may all be similar
for a given gas density, despite the presence of ``noise'' which gives
rise to minibursts.  It is important to understand the disk
self-regulation mechanism.

\end{itemize}

\section{Conclusion}

Recent progress in understanding starbursts and placing them in the
context of galaxy evolution has been outstanding and is healthily
accelerating.  We are fortunate to be riding a tidal wave of marvelous
new data highlighted by unprecedented large sample sky surveys, HST
high resolution imaging, sensitive new infrared and sub-mm
instrumentation, and the inauguration of the Spitzer 
and GALEX  observatories.

To close, let me mention some critical aspects of starburst physics
that deserve special attention. 
How does feedback operate in young starbursts to regulate processes
like saturation, quenching, and outflows?  In particular regarding the
latter, the largest effects of starbursts are related to galactic
superwinds, but there are numerous uncertainties regarding their
underlying physics.  Nearby systems are the benchmarks for detailed
scrutiny of superwinds.  A crucial open question is the mechanism
of starburst triggers: for a given trigger, there is apparently
a large variation in the resulting star-formation amplitude, which
remains poorly understood as yet.  A final important problem concerns
the drivers and time-scales for dust shroud dissipation, which
transforms an IR-bright galaxy into a UV/optical-bright one.  All
these areas will benefit from a combined observational/theoretical
attack.

\vspace{0.5cm}

All the conference participants wanted to extend their genuine
gratitude to the organizing committees for a very productive and
enjoyable conference, but most couldn't do so because of page limits.
This work has been supported in part by HST grants GO-09117 and
GO-09455.


\end{document}